# Instability patterns, wakes and topological limit sets.

R. M. Kiehn
Mazan, France
Rkiehn2352@aol.com
http://www.cartan.pair.com

**Abstract**
Many hydrodynamic instability patterns can be put into correspondence with a subset of characteristic surfaces of tangential discontinuities. These topological limits sets to systems of hyperbolic PDE's are locally unstable, but a certain subset associated with minimal surfaces are globally stabilized, persistent and non-dissipative. Sections of these surfaces are the spiral scrolls so often observed in hydrodynamic wakes. This method of wake production does not depend explicitly upon viscosity.

**Introduction**
From the topological point of view it is remarkable how often flow instabilities and wakes take on one or another of two basic scroll patterns. The first scroll pattern is epitomized by the Kelvin-Helmholtz instability (Figure 1a) and the second scroll pattern is epitomized by the Raleigh-Taylor instability (Figure 1b).

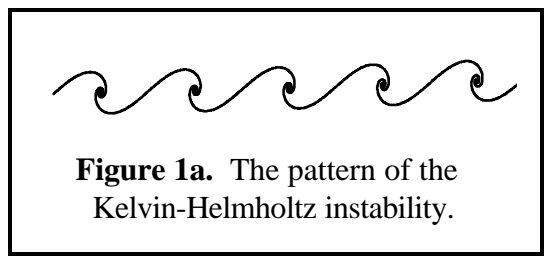

**Figure 1a.** The pattern of the Kelvin-Helmholtz instability.

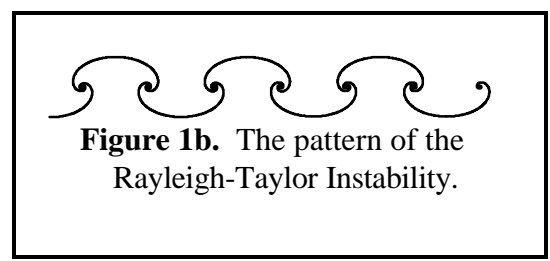

**Figure 1b.** The pattern of the Rayleigh-Taylor Instability.

The repeated occurrence of these two patterns (one similar to a Cornu Spiral and the other similar to a Mushroom Spiral) often deformed but still recognizable and persistent, even in dissipative media, suggests that a basic simple underlying topological principle is responsible for their creation. The mushroom pattern is of particular interest to this author, who long ago was fascinated by the long lived ionized ring that persists in the mushroom cloud of an atomic explosion. Although the mushroom pattern appears in many diverse physical systems (in the Frank-Reed source of crystal growth, in the scroll patterns generated in excitable systems, in the generation of the wake behind an aircraft, ...), no simple functional description of the mushroom pattern is given in the literature known to this author. Classical geometric analysis applied to the equations of hydrodynamics has failed to give a satisfactory description of these persistent structures, so often observed in many different situations.

**Viscosity vs. Compressibility**

Conventional analysis of wake production and boundary layer formation subsumes that the dominant physical effect is governed by viscous creation (and destruction) of vorticity. To quote Batchelor, "The term wake is commonly applied to the whole region of non-zero vorticity on the downstream side of a body in an otherwise uniform stream of fluid." The conventional perspective presumes that the wake is approximated as a tangential discontinuity, defined as a vortex sheet. The viscous creation of the vortex sheet is not precisely defined, but once the sheet is formed, its evolution is presumed to be governed by an integral form of the Biot-Savart law, known as the Rott-Birchoff equation. These assumptions have been shown to lead to the production of discontinuities in finite time. Asymptotic spiral type solutions in the vicinity of the singularity have been investigated both analytically and numerically [Kaden, Rott, Kambe, Pullin, Cayflisch, Kransky].

The viscous process essentially is one of diffusion, and is governed by a non-hyperbolic system of PDE's. It is most certainly applicable to the decay of wake phenomena. However, this point of view, which has its domain of applicability in the infinitely far field behind a body, is often at odds with experiments of the near field, which indicate that wake features persist with sharp definition for long periods of time, over many characteristic lengths, without diffusive blurring, and at high Reynolds numbers where viscosity effects are minimal. To quote Browand [Browand, 1986] "At the present time, there is no theoretical framework to describe the structural features of high Reynolds number shear flows."

In this article an alternate approach to the creation of wakes is proposed. The basic physical mechanism for wake production is assumed to be associated with the fact that all real fluids have a finite speed of sound, hence a finite compressibility. Therefore, there can exist domains in every (perhaps slightly) compressible fluid where the system of PDE's describing the fluid evolution is hyperbolic, and not diffusively parabolic or elliptic. Hyperbolic domains have the feature that they can be associated with topological limit sets upon which the solutions to the PDE's are not unique. Therefore topological discontinuities are admissible in such dynamical systems. Such discontinuities are of two types: shock waves, where $C^0$ differentiability is lost for the component of velocity normal to the discontinuity surface (but the tangential components of velocity are continuous) and tangential discontinuities, where $C^0$ differentiability is lost for the tangential components of the velocity, but the normal component is continuous. It is this set of limit points associated with the tangential discontinuities that can be put into correspondence with the two basic instability patterns described above. Recall that propagating discontinuities are associated with singular solutions to wave equations. In fact, the very definition of a wave, according to Hadamard, is a propagating discontinuity. To recap, the point of view taken in this article is that the *creation* of a wake is to be associated with a discontinuous process in a hyperbolic domain, while the *decay* of a wake is associated with a diffusive process in an elliptic domain.

This alternate topological approach to the problem of wake creation is independent from viscosity, and gains credence from the fact that not only is a mechanism offered for the creation of the tangential discontinuity, but also closed form solutions to the vector fields that describe the aforementioned instability patterns can be obtained. An equivalent result is not known in the literature familiar to this author. Not only does the topological point of view give new insight into how wakes may be generated, but also points out how such wakes possibly may be controlled. As described below, the resultant wake creation phenomena is closely related to the problems of diffraction and interference in electromagnetism. In fact, it may be said that from the topological point of view the near field hydrodynamic wake is a diffraction pattern caused by the physical obstruction. Hence, it is plausible that phase interference mechanisms may be used to control wakes. At high Reynolds numbers the effect of viscous diffusion is to smear out or thicken the tangential discontinuity.

**Spiral Space Curves**

The clue that wave fronts, representing propagating tangential discontinuities, are the basis for the two basic spiral instability patterns in fluid dynamics came to this author during a study of Cartan's methods of differential topology as applied to the production of defects and topological torsion in dynamical systems [ Kiehn 1990, 1991, 1992]. Cartan's ideas start from a generalization of the Frenet-Serret concept of the "moving frame". The Frenet analysis indicates that a space curve, conventionally defined by a position vector, **R**(t), relative to a specific coordinate system, is completely characterized by its arc length, s, its Frenet curvature, $\kappa(s)$, and its Frenet torsion, $\tau(s)$. [Struik 1956] All curves having the same $\kappa(s)$ and $\tau(s)$ are congruent, hence are independent from the choice of coordinates used to define the original position vector. The variables $\{s,\kappa,\tau\}$ are the intrinsic variables of a space curve.

The null set of a function $f(s,\kappa,\tau)$ defines a surface in the space of the intrinsic coordinates, and the intersection of two such surfaces defines a curve in the space of intrinsic coordinates that has a preimage in the space of initial variables, $\{x,y,z\}$. Consider "plane" curves given by the relations

$$f_1(s,\tau,\kappa) = \tau = 0 \quad \text{and} \quad f_2(s,\tau,\kappa) = \kappa - g(s) = 0 . \quad (1)\,(2)$$

For $g(s) = 0$, the intrinsic curve has zero curvature and zero torsion. It is a straight line in $\{x,y,z\}$ space, too. For $g(s) = +1$, the curve in $\{s,\kappa,\tau\}$ space is a straight line displaced from the origin to the right, parallel to the s axis and in the $\tau = 0$ plane. The preimage of this intrinsic curve with constant curvature and zero torsion in $\{x,y,z\}$ space is a circle.

For $g(s) = 1/s$ in the right half plane, the resulting space curve is the logarithmic spiral in the space $\{x,y,z\}$. For $g(s) = s$, a straight line in the intrinsic coordinate space, $\{s,\kappa,\tau\}$, the resulting $\{x,y,z\}$ image is the Cornu spiral. These facts have been known for more than 100 years to differential geometers.

However, a simple sequence is to be recognized:

$$... \kappa = g(s) = s^{-1}, \quad \kappa = s^0, \quad \kappa = s^1 ...$$

The question arises: what are the characteristic shapes of intrinsic space curves for which the curvature is proportional to an arbitrary power of the arc length, $\kappa = g(s) = s^n$, for all positive and negative integers (or rational fractions)? Through the power of the PC these questions may be answered quickly by integrating the equations,

$$dx/ds = u = \sin(Q(s)) \quad \text{and} \quad dy/ds = v = \cos(Q(s)), \qquad (3)\ (4)$$

where Q is some "phase" function of s. The vector with components, u and v, is a unit tangent vector to a curve with arc length, s. The Frenet curvature is given by $\kappa = dQ/ds$. Hence if Q is $s^{n+1}/n+1$, then the Frenet curvature is $\kappa = s^n$. The results of the numerical integrations are presented in Figure 2 for $n = 1$ and $n = 2$. A most remarkable result is that the Cornu spiral of Figure 2a is the deformable equivalent for all odd-integer $n > 0$, and the Mushroom spiral of Figure 2b is the deformable equivalent for all even-integer $n > 0$.

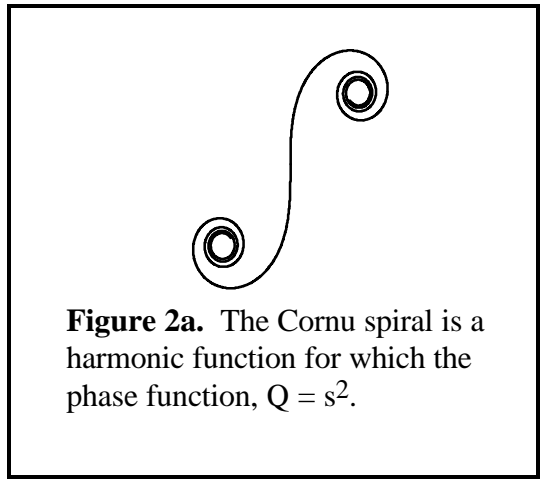

**Figure 2a.** The Cornu spiral is a harmonic function for which the phase function, $Q = s^2$.

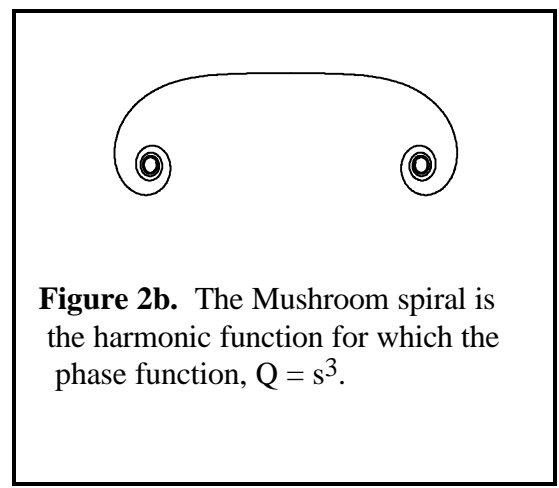

**Figure 2b.** The Mushroom spiral is the harmonic function for which the phase function, $Q = s^3$.

Not only has the missing analytic description of the mushroom spiral been found, but also a *raison d'être* has been established for the universality of the two spiral patterns. They belong to the even and odd classes of arc length exponents describing plane curves in terms of the formula, $\kappa = s^n$. The Cornu spiral is the first odd harmonic function that maps the infinite interval into a bounded region of the plane, and the mushroom spiral is the first even function that maps the infinite interval into a bounded region of the plane. Note that the zeroth harmonic function maps the infinite interval into the bounded region of the plane, but the trajectory is unique in that it is not only bounded but it is closed; i.e., a circle. A similar sequence can be generated for the half-integer exponents with n = +3/2, 7/2, 11/2... giving the Mushroom spirals and 5/2, 9/2, 13/2... giving the Cornu spirals. In numeric and experimental studies of certain shear flows, the Cornu spiral appears to dominate the motion in the longitudinal direction, while the Mushroom spiral appears in the transverse direction. Periodic patterns can be obtained by examining phase functions of various forms. For example, the Kelvin Helmholtz instability of Figure 1a is homeomorphic to the choice $Q(s) = 1/\cos^2(s)$. Similarly, the Rayleigh-Taylor instability of Figure 1b is homeomorphic to the case $Q(s) = \tan(s)/\cos(s)$.

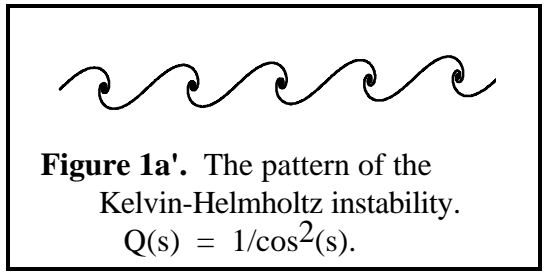

**Figure 1a'.** The pattern of the Kelvin-Helmholtz instability. $Q(s) = 1/\cos^2(s)$.

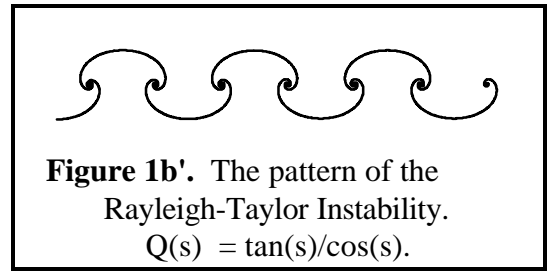

**Figure 1b'.** The pattern of the Rayleigh-Taylor Instability. $Q(s) = \tan(s)/\cos(s)$.

The Kelvin-Helmoltz instability function is highly singular at points where $\cos(s) = 0$. To study the effect of this singularity, consider a modification of the phase function, such that $Q(s) = 1/(a^2 + \cos^2(s))$. The constant term a represents the radius of a finite "core" or

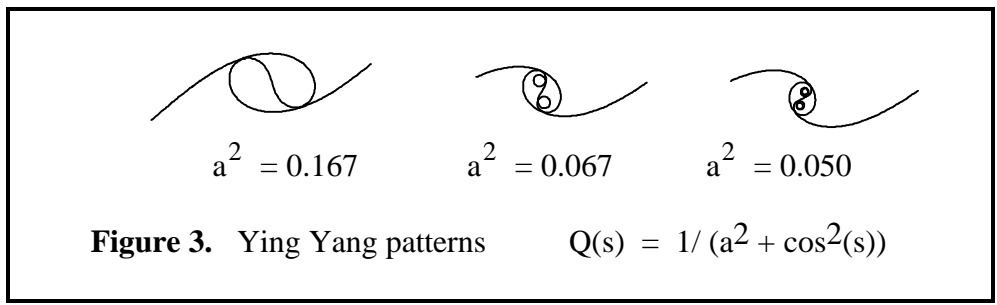

$a^2 = 0.167 \qquad a^2 = 0.067 \qquad a^2 = 0.050$

**Figure 3.** Ying Yang patterns $\qquad Q(s) = 1/(a^2 + \cos^2(s))$

hole. The solutions to (3-4) for such a phase function are extraordinary. Above a certain minimum core size the wave pattern exhibits no spiral shapes or self-intersections. At a value of $a^2 = .167$, the wave pattern describes a curve which exhibits the first double tangency, in the shape of a Ying-Yang diagram. As the core radius is decreased further, another size is reached when the double tangency again occurs, but now the pattern contains a dipole, or circulation pair. The process can be continued to generate a sequence of such double tangencies, ultimately generating the spiral patterns of Figure 1a. In the interior of each of the double tangencies is a circulation pair, about which the curve describing the pattern winds, first in one direction, then in the other direction. At each new double tangency the winding number about each doublet center increases by one. A description of this effect is given in Figure 3, where the first few double tangencies are displayed. The interpretation and application of this spin-pair production mechanism has not been developed.

**Torsion vs. Curvature**

The full Frenet equations in three dimensions are given by the expressions,

$$d\mathbf{t}/ds = +\kappa \mathbf{n}$$
$$d\mathbf{n}/ds = -\kappa \mathbf{t} + \tau \mathbf{b}$$
$$d\mathbf{b}/ds = -\tau \mathbf{n} .$$

The vectors $\{\mathbf{t}, \mathbf{n}, \mathbf{b}\}$ form an orthogonal moving frame at the point p along the space curve. The variables s, $\kappa$, $\tau$, are the arc length, curvature and torsion, respectively.

There are two extreme situations. In the first extremal case the torsion is negligible, and the space curve is confined to the plane described by the vectors $\mathbf{t}$ and $\mathbf{n}$. In classical mechanics this result is interpreted as the conservation of angular momentum. The unit binormal maintains a constant direction ($d\mathbf{b}/ds \approx 0$) along the angular momentum vector. Space curves are generated by a parametric displacement in the direction of the tangent vector. The tangent and the normal vectors form a harmonically conjugate pair if the curvature is a constant, but if the curvature is exactly zero, then the directions of the vectors $\mathbf{n}$ and $\mathbf{b}$ are indeterminate.

However, suppose that the curvature, $\kappa$, is small but finite at some initial condition, such that the normal and the binormal field can be specified. Now, however, assert that the torsion of the curve is not negligible. Then the normal and binormal form a harmonically conjugate pair. To first order, the unit tangent vector is an invariant ($d\mathbf{t}/ds \approx 0$). A curve can be generated in the $(\mathbf{n}, \mathbf{b})$ plane, parametrized by a displacement, $\nu$, along $\mathbf{b}$. If the torsion, $\tau(\nu)$, is one of the functions described above, then spiral shapes for the curve in the $(\mathbf{n}, \mathbf{b})$ plane are to be expected. Hence there are two extremes, the first extreme where the torsion tends to zero, and the space curves are dominated by curvature, and can be embedded in scrolled surfaces which locally approximate a cylinder with generators along the binormal. The second extreme is where the space curves are dominated by torsion, and are sections of a scrolled surface with local "cylindrical" generators along the tangent field. In both cases, the tangent plane to the surface is constructed (only approximately) in terms of the unit tangent and the binormal. In the first extreme, angular momentum is conserved, $d\mathbf{b}/ds \approx 0$. In the second extreme, $d\mathbf{t}/ds \approx 0$. As derived below, the curvature, $\kappa$, for an Eulerian flow is proportional to the component of the pressure gradient along the normal field for a single surface.  In certain situations, it would appear that the free wake consists of a thin double layer.

From experimental photographs, it would appear that the Kelvin-Helmholtz instability is related to characteristic spirals along the tangent line (and is curvature dominated) and the Rayleigh-Taylor instability is related to characteristic spirals along the binormal line (and is torsion dominated).  The latter case is that associated with the rollup of the wake behind and aircraft in flight, schematically described in Figure 4.   (The curvature dominated case is described in Figure 8 .)

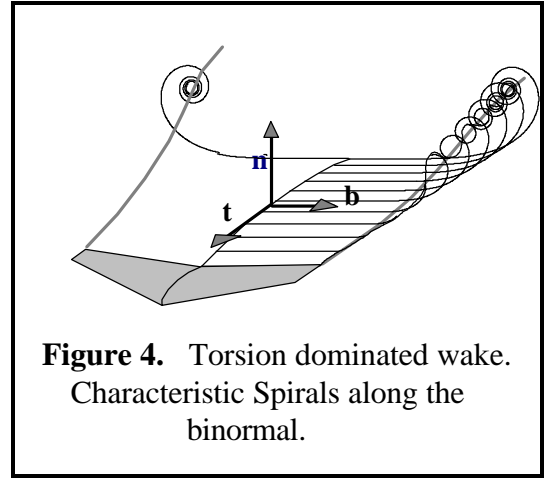

**Figure 4.**  Torsion dominated wake. Characteristic Spirals along the binormal.

**Wakes as Characteristic Minimal Surfaces**

A fundamental observation is that these spiral space curves are universally among the "characteristic" solutions to hyperbolic systems of PDE's, of which the Euler equation of two dimensional compressible flow,

$$A\phi_{\eta\eta} + 2B\phi_{\eta\xi} + C\phi_{\xi\xi} = D,$$

is a prime example. Following Landau [Landau, 1959, p.380], the characteristic surfaces contain embedded curves which are given by solutions to ordinary differential equations,

$$d\eta/d\xi = [B \pm (B^2 - AC)^{1/2}]/C.$$

Recall that characteristic surfaces represent point sets upon which the solutions to the PDE's are not unique; i.e., the characteristic surfaces include surfaces of tangential discontinuities (on which the normal components of the vector field are continuous, but the tangential components are not continuous), and shocks (on which the normal components are discontinuous, but the tangential components are continuous).  Shocks permit the flow of mass across the discontinuity and permit pressure discontinuities, tangential discontinuities do not.  Shocks can be dissipative and involve entropy change; tangential discontinuities are not dissipative.  Shocks are generally stable, tangential discontinuities are generally unstable.  Landau states that tangential discontinuities are to be ignored, for they will lead to turbulence.   However, a certain subset of characteristic surfaces, which always are locally unstable, may be globally stabilized. It is these special characteristic surfaces, those associated with minimal surfaces,  that are of that interest to this article.

Define a position vector {u,v,w} to a point on a surface in terms of the characteristic coordinates, with parametrization, $\sigma$, as

$$u = d\eta/d\sigma = A(\sigma) \sin(Q(\sigma)),$$
$$v = d\xi/d\sigma = A(\sigma) \cos(Q(\sigma)),$$
and $\quad w = F(u,v) = f(\sigma).$ (5)

If $F(u,v)$ satisfies the equation

$$(1 + F_v^2)F_{uu} - 2 F_u F_v F_{uv} + (1+ F_u^2)F_{vv} = 0 \qquad (6)$$

then $F(u,v)$ defines a minimal surface. There is one solution of (6) which is unique in that it is the only *harmonic* minimal surface; it is the solution $F(u,v) = \tan^{-1}(u/v) = Q(\sigma)$, or the right helicoid [Osserman, 1983]. It is this special subset of characteristic surfaces that is to be associated with the spiral solutions generated by Equations (3-4). Working backwards, assert that the surface of characteristics is a minimal surface; then to generate the ordinary differential equations (3-4), the minimal surface must be the right helicoid. More precisely, the fundamental result is that of the infinite number of surfaces of tangential discontinuities, there is a special subset whose sections yield spiral space curves that are solutions of Equations (3-4). It turns out that this special subset can be related to the unique harmonically generated minimal surface, the right helicoid.

As mentioned above, all such surfaces of tangential discontinuities are locally unstable [Landau, 1959 p.114], for they are associated with a hyperbolic domains. In fact Landau claims that these surfaces are the precursors of turbulence. However, the special subset of locally unstable surfaces of tangential discontinuities (in fact those which contain the spiral characteristic lines described above, and which can be associated with minimal surfaces) can, like soap films, exhibit domains of global stability [Barbossa, 1976]. In the domain of global stability, the locally unstable surfaces create the persistent wake patterns observed at moderate Reynolds numbers. When the Reynolds number exceeds a certain value such that the global stability of the minimal surface is lost, then these special surfaces of tangential discontinuity lose their global stability and the flow becomes turbulent. It is the thesis of this article that it is these globally stabilized surfaces of tangential discontinuities, which are not diffusive, but which are persist and observable, which are those surfaces that generate the distorted but persistent mushroom patterns in the Von Karman wake, or the Lanchester tip vortices in the wake of an aircraft in flight, or the Kelvin-Helmholtz instability pattern of a shear layer.

Such patterns are reproduced in a qualitative way by appropriate adaptations of the phase function $Q(s)$. For example, in Figure 5a, the phase function used to generate the Kelvin-Helmoltz instability of Figure 1a has been modified to include an amplitude factor $A(s) = .1 \, s^{3/2}$. The exponent of the amplitude factor is related to the ratio of the relative scales between successive spiral singularities. The physical interpretation of this mathematical observation has not been resolved, but it points to an experimental property that could be measured. If the phase function used to describe the Rayleigh-Taylor instability of Figure 1b is modified to include a translation term proportional to cos(s), then the spiral wave pattern of Figure 5b is produced. It is apparent that this modification of the phase function is related to the staggered Von Karman wake.

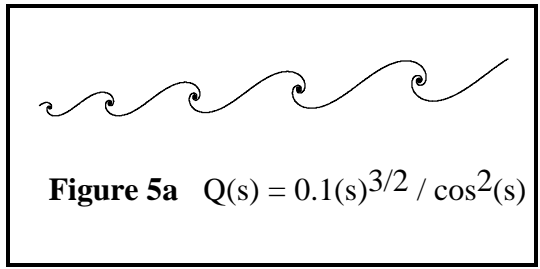
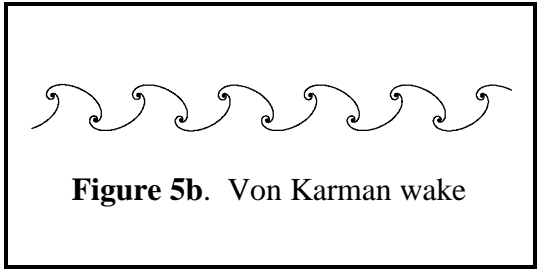

**Figure 5a**  $Q(s) = 0.1(s)^{3/2} / \cos^2(s)$

**Figure 5b**. Von Karman wake

Further modifications of the Von Karman wake of Figure 5b to include a decaying minimal core radius produces the complex spiral pattern of Figure 6. If the phase factor includes a linear and a quadratic term in arc length s, as well as the Kelvin-Helmholtz term, then the generated pattern imitates the observed flow pattern generated around a sharp edge. See Figure 7.

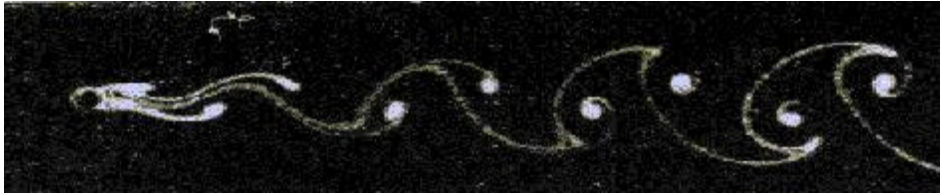
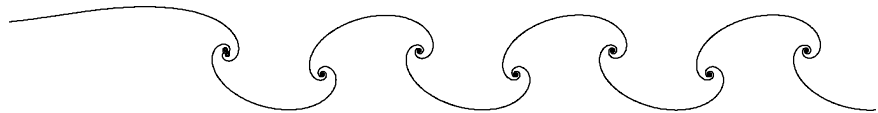

**Figure 6.** Von Karman Wake past a cylinder.

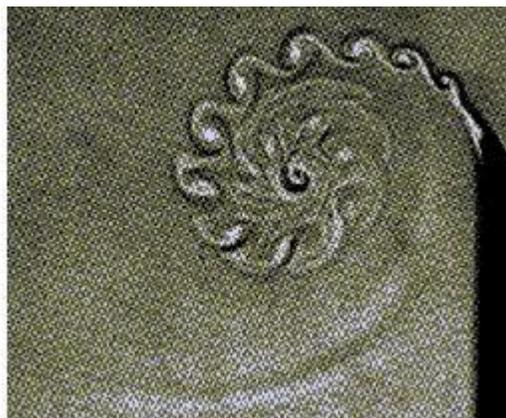
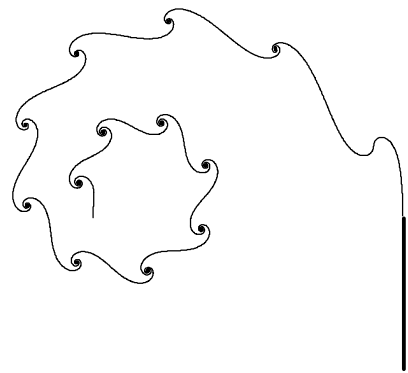

**Figure 7.** Wake past a sharp edge

The fact that the spiral characteristic lines are contained in the only harmonic minimal surface (generated by the function w = arctan Q(s)), and their obvious connection with the Cornu Spiral, demonstrates that the production of the Kelvin-Helmholtz instability pattern is a first order (odd) diffraction problem. It seems natural to suggest that the Rayleigh-Taylor instability pattern is a second order (even) diffraction problem. Being topological limit sets, these sets of tangential discontinuities can create extensions of material boundaries into the bulk fluid. Typically, in the region of a "sharp" edge, a material boundary will be extended by a topological limit set of tangential discontinuities in the form of a scroll pattern, independent from any viscous, diffusive effect. During the transient accelerations from rest to a state of constant velocity, the flow about sharp edges will become hyperbolic, and (in section) lines of tangential discontinuity will grow into the bulk fluid. These surfaces (in section, lines) of topological limit sets form extensions of the material boundary, and will evolve into the spiral shapes described above. As mentioned above, tangential discontinuities also can be generated at points where the pressure thermodynamically permits a change of phase, or cavitation, to take place. For example, in potential flow around a cylinder, the kinetic energy density at the top of the cylinder is four times the kinetic energy density at infinity. At points along the streamline in the neighborhood of the cylindrical surface, between the leading stagnation point and the top of the cylinder, and where the kinetic energy density exceeds one third the pressure energy density at infinity, Bernoulli's law would indicate a domain of "zero" or "negative" pressure. Physically, it would be expected that a surface of tangential discontinuity (separation) would be created when the Bernoulli pressure goes to zero.

The characteristic surface is a domain where more than one solution to a set of PDE's can exist. In the hydrodynamic case, such a surface can be used to divide a domain where the flow in the nearby neighborhood is represented by a non-harmonic potential function (div $\mathbf{V} = \nabla^2\phi \neq 0$, curl $\mathbf{V} = 0$) from a nearby neighborhood where the flow is represented by a non-harmonic stream function ( div $\mathbf{V} = 0$, curl $\mathbf{V} = \nabla^2\psi \neq 0$). In the first case, the fluid is compressible, but without vorticity, and in the second case the fluid is incompressible, but with vorticity. A classical wake could be modeled as a thin layer bounded on either side by such a surface of tangential discontinuity. On the surfaces of boundary limit points, both functions are harmonic, but admit discontinuities. The interior of the thin layer would be diffusive, and would exhibit a evolutionary thickening. The tangential discontinuity would not permit mass flow between the two regions. In the classic analysis of wakes, the viscous-vorticity point of view (div $\mathbf{V} = 0$, curl $\mathbf{V} = \nabla^2\psi \neq 0$) has dominated the literature. In this article, the alternate point of view based upon compressibility (div $\mathbf{V} = \nabla^2\phi \neq 0$, curl $\mathbf{V} = 0$) is assumed to be dominant, for it gives not only a plausible physical theory for describing the production of the tangential discontinuity, but also leads the way to finding close form generators for the wake patterns created by such tangential discontinuities.

**Integrability and Topological Defects**

The two extremes of wake patterns mentioned above indicate that the vector field tangent to the spiral characteristics is (to first order) two dimensional. More precisely, the observation is that these characteristic lines (not necessarily the flow field) are integrable in the sense of Frobenious. Either the tangent field is integrable, or the binormal field is integrable. It is of some interest to see how these two extremes for the characteristics are related to hydrodynamic flows. To this end, a useful procedure is obtained by making a transformation to a space-time coordinate system involving the intrinsic properties of the particle paths (in this case streamlines). That is, consider the velocity field, **V**, which may be composed of a unit tangent field, **t**, and a normalization factor U.

$$\mathbf{V} = U\,\mathbf{t}.$$

Assume that the unit tangent field is not explicitly time dependent (such that the particle paths are along streamlines), but the normalization factor might be time dependent, U = U(x,y,z,t).

This representation for a vector field is very convenient, for if a unit tangent field can be found that satisfies boundary conditions at infinity and has zero normal component along a material boundary, U(x,y,z,t) = 0, then the zero set of the product function, **V** = U(x,y,z,t) **t**(x,y,z), can be used to define no-slip boundary conditions. The unit tangent field, **t**, is permitted to have its own singularities (zeros), which are different from the zero sets of the normalization function, U. On the surface U(x,y,z,t) = 0, the velocity field vanishes identically.

The equations for the divergence and curl of the vector field become

$$\mathbf{V} = U\,\mathbf{t}, \qquad\qquad \mathbf{V}\bullet\mathbf{V}/2 = U^2/2,$$
$$\mathrm{div}\,\mathbf{V} = \mathrm{grad}U \bullet \mathbf{t} + U\,\mathrm{div}\,\mathbf{t}, \quad \mathrm{curl}\,\mathbf{V} = \mathrm{grad}U \times \mathbf{t} + U\,\mathrm{curl}\,\mathbf{t},$$
$$\mathbf{V} \bullet \mathrm{curl}\,\mathbf{V} = U^2\,\mathbf{t} \bullet \mathrm{curl}\,\mathbf{t},$$
$$\mathbf{V} \times \mathrm{curl}\,\mathbf{V} = U\,\mathrm{grad}U - U\,(\mathbf{t} \bullet \mathrm{grad}U)\,\mathbf{t} + U^2\,\mathbf{t} \times \mathrm{curl}\,\mathbf{t},$$

The divergence in the nearby vicinity of the set U(x,y,z,t) = 0 is related to gradU • **t**, and the curl is equal to gradU x **t**. Note that curl **V** always resides in the surface U(x,y,z,t) = 0, (i.e., the vorticity is in the tangent plane of the surface whose normal is given by grad U). The divergence, div **V**, is not zero unless **t** resides in the surface represented by equation, U(x,y,z,t) = 0.

The non-exact differential, ds, of arc length is defined such that

$$d\mathbf{R} = \mathbf{t}\,ds = \mathbf{V}\,dt\,.$$

Assuming locally a euclidean metric, it follows that ds = U dt. The differential dt is exact; the differential ds is not. Closed loop integrals for dt vanish, but closed loop integrals for ds do not. From the Frenet theory,

$$d\mathbf{t}/ds = \kappa \mathbf{n} = (\partial \mathbf{t}/\partial t + \partial \mathbf{t}/\partial x^\upsilon \, dx^\upsilon/dt)(1/U) = (1/U)\partial \mathbf{t}/\partial t + \operatorname{curl} \mathbf{t} \times \mathbf{t}. \quad (7)$$

where $\kappa$ is the curvature, $\mathbf{n}$ is the normal vector, orthogonal to the unit tangent field. Curl $\mathbf{t}$ is the Darboux vector,

$$\operatorname{curl} \mathbf{t} = \tau \mathbf{t} + \kappa \mathbf{b}, \quad (8)$$

given in terms of the binormal, $\mathbf{b}$, of the Frenet frame, and the torsion, $\tau$, of the space curve, as well as the curvature, $\kappa$, and the unit tangent field, $\mathbf{t}$.

The Frenet theory becomes useful in a hydrodynamic representation when it is realized the convective derivative may be expressed in terms of the Frenet derivative:

$$\begin{aligned} d\mathbf{V}/ds &= (d\mathbf{V}/dt)dt/ds \\ &= (\partial \mathbf{V}/\partial t + \partial \mathbf{V}/\partial x^\upsilon \, dx^\upsilon/dt)(1/U) \\ &= (\partial \mathbf{V}/\partial t + \mathbf{V} \bullet \operatorname{grad} \mathbf{V})(1/U) \\ &= (\partial \mathbf{V}/\partial t + \operatorname{grad}(U^2/2) - \mathbf{V} \times \operatorname{curl} \mathbf{V})(1/U) \\ &= (dU/ds)\mathbf{t} + U \operatorname{curl} \mathbf{t} \times \mathbf{t} + \partial \mathbf{t}/\partial t \\ &= (dU/ds)\mathbf{t} + U \kappa \mathbf{n} + \partial \mathbf{t}/\partial t. \end{aligned}$$

The Euler equations of hydrodynamics therefore have an intrinsic representation as

$$(dU/ds)\mathbf{t} + U \, d\mathbf{t}/ds = (dU/ds)\mathbf{t} + U\kappa \mathbf{n} + \partial \mathbf{t}/\partial t = -\operatorname{grad} P/(\rho U) + \mathbf{f}/(\rho U), \quad (9)$$

or,

$$[(\rho U)dU/ds + \mathbf{t} \bullet \operatorname{grad} P]\mathbf{t} + [(\rho U^2 \kappa + \mathbf{n} \bullet \operatorname{grad} P]\mathbf{n} = \mathbf{f} - [\mathbf{b} \bullet \operatorname{grad} P]\mathbf{b}. \quad (10)$$

If the external forces were identically zero, it would follow that the square bracket factors must vanish everywhere, and the curvature or bending becomes

$$\kappa = -\mathbf{n} \bullet \operatorname{grad} P / \rho U^2. \quad (11)$$

Pressure gradients in the direction of the binormal cannot be sustained without a component of external force in he same direction. In other words, the Euler equations (with no external forces) constrain the flow such that the torsion is null and the curvature is given by (11)

Equation (10) is predicated upon the assumption that the velocity vector field can be represented by a single parameter group of evolution, such that $d\mathbf{R} - \mathbf{V}dt = 0$. Suppose, however that this is only true on an "equilibrium singular" surface, in the sense that off this surface, $d\mathbf{R} - \mathbf{V}dt = \mathbf{e}$, where $\mathbf{e}$ is a small vector of fluctuations. The external forces, $\mathbf{f}$, then could be attributed to fluctuations. In any case rewrite (10) as:

$$A(s)\mathbf{t} + B(s)\mathbf{n} = (A^2 + B^2)^{1/2}d\mathbf{P}/ds = \mathbf{f} - C(s)\mathbf{b}. \quad (12)$$

The zero set of the norm of this vector defines the equilibrium singular surface. Construct the derivative of this constraint equation with respect to s to yield,

$$(dA/ds - \kappa B)\,\mathbf{t} + (dB/ds + \kappa A - \tau C)\,\mathbf{n} + (dC/ds + \tau B)\,\mathbf{b} = d\mathbf{f}/ds. \qquad (13)$$

First hypothecate the vector **f** in the direction of the binormal is vanishingly small. Then, subject to the assumptions imposed above, $C = 0$, and $dC/ds = 0$, and the torsion, $\tau$, must vanish:

$$\tau = \mathbf{t} \bullet \operatorname{curl} \mathbf{t} = \mathbf{V} \bullet \operatorname{curl} \mathbf{V}/U^2 = 0. \qquad (14)$$

This condition is equivalent to the Frobenius integrability condition. Note that the integrability condition is equivalent to the vanishing of the helicity density, $\mathbf{V} \bullet \operatorname{curl} \mathbf{V} = 0$, for the flow. The flow field is of Pfaff dimension 2 or less. The coefficients A and B are *harmonic* in the sense that,

$$dA/ds - \kappa B) = 0,$$
$$(dB/ds + \kappa A) = 0. \qquad (15)$$

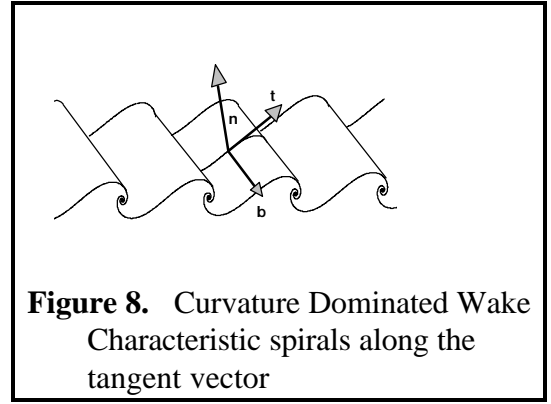

**Figure 8.** Curvature Dominated Wake Characteristic spirals along the tangent vector

In fact, divide (15) by $(A^2 + B^2)^{1/2}$ to form a unit tangent vector with the same format as equations (3-4); spiral solutions in the direction of the tangent field are to be expected for the curvature dominated characteristics of such a surface. Figure 8 is a schematic of the curvature dominated case.

Suppose that **f** is not identically zero, but $d\mathbf{f}/ds$ is approximated by zero. Then the "power" equation given by (13) can be examined in the limit of $\kappa \approx 0$. This would be the case of torsion dominated evolution as discussed above. For torsion dominated flows, the pressure gradient is either perpendicular to the normal, or zero. The spiral characteristic lines are in the direction of the binormal. A schematic representation of the torsion dominated case was presented in Figure 4. Note that the velocity field is NOT integrable in the Frobenius sense, for a torsion dominated wake. The flow field is of Pfaff dimension 3 or higher.

The topological defects (characteristic limit sets) described above are associated with flow fields of Pfaff dimension 2 (vector fields that satisfy the Frobenius integrability theorem) and Pfaff dimension 3. Other topological defects for flow fields of Pfaff dimension higher than 2 yield more complicated wakes. For example, in a rotating frame of reference it is possible to find closed form solutions to the Navier-Stokes equations which are of Pfaff dimension 3. Such vector fields have finite torsion, $\tau$, and exhibit re-entrant flow patterns (contained within a swirling bubble) above a critical flow parameter.

The topological defect associated with the transition to a re-entrant flow is not a dislocation hole, but can be put into correspondence with when the surface of zero torsion changes from a hyperboloid of one sheet into a hyperboloid of two sheets. The surface of zero torsion becomes *disconnected*. The analytic results, which mimic the "vortex" bursting problem, are presented in [Kiehn, 1991]. Note that in this problem the sheet of zero torsion is a point set (a surface, not flat) upon which the Frobenius integrability conditions are true.

**Lagrangian Wakes and Dislocation Defects**

When a harmonic potential flow or a harmonic stream function is used to describe a fluid flow, the symmetry of the Eulerian streamlines created about obstacles gives no indication of a wake. However, if it is assumed possible to synchronize globally and transversely a set of "points" in a fluid, then the kinematic evolution of such a set will give an indication of a wake. In this section these ideas will be examined. Potential flows (irrotational vector fields) are always sychronizable over the domain of uniqueness. Stream function flows (as solenoidal vector fields constructed from two independent variables at most, in 3 dimensions) are also globally synchronizable over the domain of uniqueness. However, the combination of a potential flow and a solenoidal flow in 3D is not necessarily integrable

The Frobenius integrability theorem can be used to determine if a given vector field admits globally a set of points that can be synchronized. That is, if a set of initial points lies on a smooth surface everywhere transversal to the flow field, then after an interval of evolution (time increment) the initial set of points can still be connected by a smooth surface in the final state if and only if the vector field satisfies the Frobenius condition. This fact implies that the Pfaff dimension of the vector field never exceeds 2 over the domain of interest, and that there exists a global submersion of this domain to a space of two dimensions! The smooth surface of synchronous points may be grossly distorted, and this distortion is indicative of different topological features of the flow and its wake.

For example, consider the two dimensional potential flow about a cylindrical rod. Start a set of synchronized particles at very far distances upstream from the obstacle. Assert that at the initial position the synchronized set of points resides on a surface that is orthogonal to the streamlines, much like a plane wave front in optics. Now forward integrate the kinematic equations for these points using the velocity potential for the streamline flow. As the "wave front" advances to the obstacle, the front retains its shape as a plane which is transversal and almost orthogonal to the streamlines. However as the "wave front" reaches the obstacle, the forward stagnation point greatly deforms the shape of the "synchronized" set of points that make up the "wave front". After the "wave front" of synchronous points has passed over the obstacle, a noticeable near field wake is observed downstream, a wake pattern that is not at all apparent from the format of the streamlines, which do form a pattern which is symmetric with respect to time inversion. The Eulerian description of the flow pattern is symmetric, but the Lagrangian description, with a choice of initial conditions, is not.

At very far distances to the rear of the obstacle, the "wave front" for synchronous points again looks as if were a plane surface almost perpendicular to the flow lines. Different times correspond to uniformly spaced "almost plane waves". Only very close inspection (on scales smaller that the obstacle size) reveals that in the neighborhood of the positive x-axis the set of equal time surfaces has "sharp" bend on the downstream side of the obstacle. The equal time surface does not cross the x-axis downstream from the obstacle, but it does cross the x axis on the upstream side. See Figure 9a. On measurement scales much larger than the radius of the cylindrical obstruction, the wave fronts form a lattice with only a "local" defect in the downstream side of the obstacle. This defect (on these very large scales) probably would be ignored observationally.

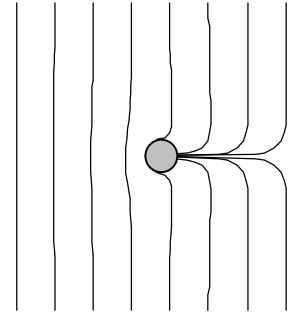

**Figure 9a**  Lagrangian wake, no circulation

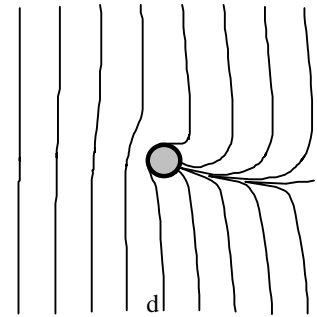

**Figure 9b.**  Lagrangian wake with circulation. Note the zebra stripe dislocation defect.

On the other hand, if the potential flow that generated the wake of Figure 9a is modified to include some circulation, but no vorticity, then the "wave front" pattern exhibits an apparent "global" defect as it passes over the obstacle. On scales much larger than the radius of the cylindrical obstacle, the "plane waves" that approximate the synchronous particles have a "dislocation" defect in the downstream wake. This topological defect, associated with the circulation of the flow about the obstacle, has nothing to do with vorticity! It would not be ignored observationally. See Figure 9b. An extra line, or defect, appears to be inserted into the large scale pattern. The circulation integral around the contour that is a section of the closed surface composed of two equal time faces is zero, but the contribution from the loop around the circular cylinder is balanced by the contribution along the dislocation defect.

This work was supported in part by the ISSO at the University of Houston.